\documentclass[a4paper,fleqn]{cas-sc}
\usepackage{lineno}
\usepackage[numbers,sort&compress]{natbib}
\usepackage{amsmath}
\usepackage{amssymb}
\usepackage{graphicx}
\usepackage{float}
\usepackage[section]{placeins}
\usepackage{booktabs}
\usepackage{tikz}
\usetikzlibrary{arrows.meta,positioning,shapes.geometric}

\graphicspath{{figs/}}

\begin{document}
\let\WriteBookmarks\relax
\renewcommand{\topfraction}{0.90}
\renewcommand{\bottomfraction}{0.80}
\renewcommand{\textfraction}{0.08}
\renewcommand{\floatpagefraction}{0.80}
\setcounter{topnumber}{4}
\setcounter{bottomnumber}{2}
\setcounter{totalnumber}{6}

\shorttitle{Beam spill structure studies at CEE}
\shortauthors{Lv et~al.}

\title[mode=title]{Beam Spill Structure Measurements at CEE Using the T0 Detector}
\tnotemark[1]

\tnotetext[1]{This work was supported by the National Natural Science Foundation of China under Grant Nos. 11927901, 12205296 and 11975228, the State Key Laboratory of Particle Detection and Electronics under Grant No. SKLPDE-ZZ-202416, and the USTC Research Funds of the Double First-Class Initiative under Grant No. WK2030000052.}

\author[1,2]{Shuaike Lv}
\credit{Data curation, Formal Analysis, Writing - Original draft preparation}

\author[1,2]{D. Hu}[type=editor,
orcid=0000-0002-5749-9047]
\cormark[1]
\ead{hurongr@ustc.edu.cn}

\credit{Conceptualization of this study, Methodology, Software}

\author[3]{Dayu Yin}
\credit{Investigation, Data curation, Validation}

\author[1,2]{Haoqian Xu}
\credit{Investigation, Data curation, Validation, Writing - Review \& Editing}

\author[3]{Herun Yang}
\credit{Investigation, Data curation, Validation}

\author[3]{He Zhao}
\credit{Investigation, Data curation, Validation}

\author[1,2]{JiaJun Qin}
\credit{Methodology, Writing - Review \& Editing}

\author[1,2]{Junfeng Yang}
\credit{Investigation, Validation}

\author[3]{Juan Ruan}
\credit{Investigation, Data curation, Validation}

\author[3]{Jie Li}
\credit{Investigation, Data curation, Validation}

\author[1,2]{Ke Sun}
\credit{Investigation, Validation}

\author[1,2]{Lei Zhao}
\credit{Methodology, Writing - Review \& Editing}

\author[3]{Lijun Mao}
\credit{Investigation, Data curation, Validation, Writing - Review \& Editing}

\author[1,2]{M. Shao}[type=editor,
orcid=0000-0002-2268-5624]
\ead{swing@ustc.edu.cn}
\credit{Conceptualization of this study, Methodology, Software}

\author[3]{Min Li}
\credit{Investigation, Data curation, Validation}

\author[3]{Meitang Tang}
\credit{Investigation, Validation}

\author[1,2]{Shiya Huang}
\credit{Investigation, Formal analysis, Validation, Writing - Review \& Editing}

\author[3]{Tong Liu}
\cormark[3]
\ead{liutong@impcas.ac.cn}
\credit{Conceptualization of this study, Methodology, Software}

\author[3]{Tingting Hou}
\credit{Investigation, Data curation, Validation}

\author[3]{Weilong Li}
\credit{Investigation, Data curation, Validation}

\author[3]{Wenwen Ge}
\credit{Investigation, Data curation, Validation}

\author[3]{XiangLun Wei}
\credit{Investigation, Data curation, Validation}

\author[3]{Yuhong Yu}
\cormark[4]
\ead{yuyuhong@impcas.ac.cn}
\credit{Conceptualization of this study, Methodology, Software}

\author[1,2]{Yingjie Zhou}
\credit{Investigation, Validation, Writing - Review \& Editing}

\author[1,2]{Yiding Zhao}
\credit{Writing - Review \& Editing}

\author[3]{Yapeng Zhang}
\credit{Investigation, Data curation, Validation}

\author[3]{Zhoubo He}
\credit{Investigation, Data curation, Validation}

\author[3]{Zhijie Li}
\credit{Investigation, Data curation, Validation}

\author[3]{Zixuan He}
\credit{Investigation, Data curation, Validation}

\author[3]{Ziyang Wei}
\credit{Investigation, Data curation, Validation}

\author[3]{Zhiyu Sun}
\credit{Investigation, Writing - Review \& Editing}

\author[1,2]{Zhengyang Sun}
\credit{Investigation, Validation, Writing - Review \& Editing}


\cortext[cor1]{Principal corresponding author}
\cortext[cor2]{Corresponding author}
\cortext[cor3]{Corresponding author}
\cortext[cor4]{Corresponding author}

\affiliation[1]{organization={State Key Laboratory of Particle Detection and Electronics, University of Science and Technology of China},
             addressline={96 Jinzhai Road},
             city={Hefei},
             postcode={230026},
             state={Anhui},
             country={China}}

\affiliation[2]{organization={Department of Modern Physics, University of Science and Technology of China},
             addressline={96 Jinzhai Road},
             city={Hefei},
             postcode={230026},
             state={Anhui},
             country={China}}
\affiliation[3]{organization={Institute of Modern Physics, Chinese Academy of Sciences},
             addressline={509 Nanchang Rd},
             city={Lanzhou},
             postcode={730000},
             state={Gansu},
             country={China}}

\begin{abstract}
The Cooling Storage Ring External-target Experiment (CEE) at the Heavy Ion Research Facility in Lanzhou uses a slow-extracted heavy-ion beam whose temporal profile directly affects detector occupancy. The existing CEE T0 detector, originally designed to provide the event start time, is extended in this work to measure the spill structure immediately upstream of the fixed target. The detector consists of a 0.8-mm-thick EJ-200 plastic scintillator read out by 16 silicon photomultipliers at each end. The SiPM signals are conditioned and discriminated by the NINO front-end electronics, and the resulting timing signals are recorded using an FPGA-based time-to-digital converter. In laboratory tests, with the laser trigger set to its maximum available repetition frequency of 100~MHz, the recorded T0 hit-rate distribution was concentrated around 100~MHz. During the June 2024 beam test with 350-MeV/u Kr ions, the accelerator team varied several operating settings, including RF-KO, sextupole, and feedback settings, while T0 independently recorded the corresponding spill profiles. An empirical representation was used to extract the rise, plateau, and fall durations from the stored profiles. The plateau duration was used as the primary comparison metric while accelerator stability and RF-system load were also considered. The measurements clearly distinguished the spill-profile responses associated with different accelerator settings. The spill obtained with the selected settings exhibited a more nearly rectangular envelope than the initial profile, although residual fluctuations and a terminal spike remained. These results demonstrate that the existing CEE T0 detector can provide independent beam-structure information during accelerator tuning.

\end{abstract}

\begin{highlights}
\item The existing CEE T0 detector measures spill structure upstream of the target.
\item Laser tests give a recorded T0 hit rate consistent with the 100-MHz input.
\item T0 profiles compare RF-KO, sextupole, and feedback settings.
\item T0 measurements resolve spill-profile changes associated with accelerator settings.
\end{highlights}


\maketitle

\section{Introduction}
\label{sec1}
Beam diagnostics constitutes a cornerstone technology in accelerator-based nuclear-physics experiments. Accurate determination of the beam's temporal structure, intensity profile, and stability is essential not only for evaluating accelerator performance, but also for safeguarding the normal operation of downstream detection systems and ensuring the reliability of experimental data\cite{Forck2020}. In fixed-target experiments employing slow-extraction techniques, the time profile of the extracted beam directly defines the instantaneous counting rate that the detectors must sustain, and its uniformity is a critical factor governing data quality.

An ideal slow-extraction process should maintain an approximately constant beam intensity throughout the extraction window, thereby delivering a near-uniform event rate per unit time\cite{Yang2022}. In practice, the extracted-beam profile can exhibit substantial temporal fluctuations, sometimes accompanied by a sharp peak near the end of extraction. During the present CEE beam study, the accelerator team adjusted several machine-control settings, including the platform RF voltage, the maximum kick amplitude and central frequency labelled for RF-KO operation, the sextupole ramp, and the feedback gain. This work compares the corresponding spill profiles measured by T0 and does not attempt to determine the accelerator-physics mechanism associated with each setting. A non-uniform profile reduces beam-utilization efficiency and can expose downstream detectors to short periods of excessive occupancy.

The CEE experiment is a fixed-target heavy-ion spectrometer at the HIRFL-CSR external-target terminal and incorporates multiple detector subsystems\cite{Lu,XIA200211}. These include a T0 detector\cite{HU2023168773}, inner and external time-of-flight detectors\cite{Wang_2022,Wang_2020}, a time projection chamber (TPC), and multi-wire drift chambers. Rapid variations in the instantaneous beam intensity increase detector occupancy and can cause pile-up or dead-time losses, particularly in the TPC, thereby degrading track reconstruction and subsequent physics analysis. Precise diagnosis and optimization of the extracted-beam profile are therefore important for stable operation of the CEE experiment.

The accelerator ionization chamber provides reliable monitoring of the overall beam intensity and remains the sensor used by the extraction-feedback system. However, conventional beam diagnostic instruments, such as ionization chambers, provide stable and reliable monitoring of the average beam intensity. Their typical response time is on the order of microseconds, and in charge‑sensitive mode, the maximum counting rate is limited to approximately 1 kHz. In charge‑integration mode, they are unable to detect weak beams—for example, for a carbon beam at 1.2 GeV/u, the lower detection limit is $10^5$ Hz, whereas the beam intensity required at the CEE terminal is only $10^4$ Hz. As a result, such devices cannot resolve the rapidly varying beam structures encountered during slow extraction, limiting their utility for real‑time beam‑profile optimization. To address the need for high‑rate beam diagnostics, detector technologies with faster response and higher rate capability must be developed. For the present study, the CEE T0 detector provides a complementary event-counting measurement immediately upstream of the target. Its position and direct response to incident particles make it useful for comparing the beam profiles obtained under different extraction settings. The T0 detector and its readout chain must be sufficiently fast so that closely spaced signals can still be separated during periods of high instantaneous rate.

The CEE T0 detector uses a plastic scintillator coupled to silicon photomultipliers (SiPMs), providing a compact structure and the timing performance required for the event-start measurement. For its additional use in high-rate beam monitoring, however, the relatively long fall time of the raw SiPM signal can lead to pulse pile-up and counting losses.

To meet the beam-monitoring requirements of CEE, we use the existing T0 start-time detector with its signal-conditioned readout for high-rate beam-profile measurements. The readout incorporates a compact coupling and pulse-shaping network that shortens the SiPM waveform before it enters the NINO amplifier--discriminator\cite{Anghinolfi2004,Cao2022}, and the resulting detector hits are recorded by the DAQ. Laboratory tests show stable digital output up to the 100~MHz limit of the available laser source, providing the experimental basis for using this readout configuration in high-rate beam monitoring.

This work adds a beam-monitoring function to the existing CEE T0 detector to measure the spill structure at the target location. During beam commissioning, the accelerator team adjusted RF-KO, sextupole, and feedback settings, while T0 independently recorded the corresponding hit-rate profiles. The measured spill durations provide diagnostic information for accelerator tuning, with plateau duration used as the primary comparison metric after accounting for long-term accelerator stability and RF-system load. The paper is organized as follows. Section~\ref{sec2} describes the T0 detector, readout electronics, and pulse-shaping network. Section~\ref{sec3} presents the laboratory high-rate test. Section~\ref{sec4} presents the T0 measurements obtained during accelerator-setting scans and compares the corresponding spill-profile parameters. Finally, Section~\ref{sec5} summarizes the demonstrated performance and the remaining limitations.

\section{T0 detector and readout electronics}
\label{sec2}

\begin{figure}
\centering
\includegraphics[width=0.85\linewidth]{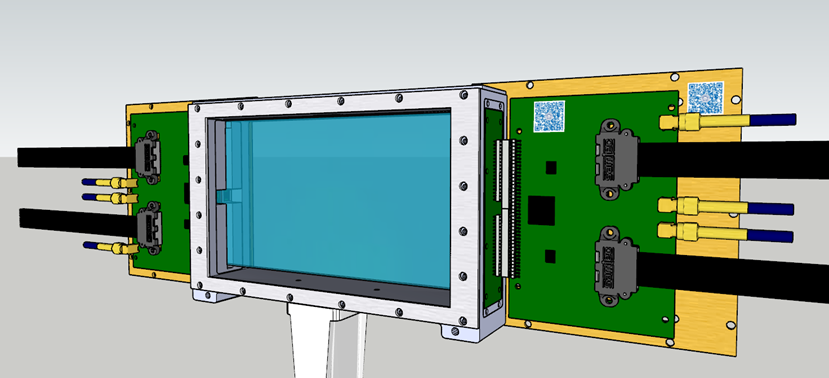}
\caption{Mechanical layout of the CEE T0 detector. The EJ-200 scintillator is read out by 16 SiPMs at each end.}
\label{fig:t0-layout}
\end{figure}

The CEE T0 detector supplies the event start time and is installed immediately upstream of the fixed target. Its sensitive element is an EJ-200 plastic-scintillator plate\cite{ej200} with dimensions \(200\,\mathrm{mm}\times100\,\mathrm{mm}\times0.8\,\mathrm{mm}\). The small thickness limits the amount of material traversed by the incident beam before it reaches the target. Depending on the species and properties of the incident beam, the scintillator thickness can be adjusted to balance the deposited energy, timing performance, and material budget. When a charged particle crosses the plate, the deposited energy produces scintillation light that propagates toward the two opposite edges. Sixteen Hamamatsu S13360-3025PE SiPMs\cite{s13360} are coupled to each edge, giving 32 independently read out optical channels. The two-sided arrangement collects light from both ends of the scintillator and provides the timing signals used to determine the event start time. The mechanical layout and detector housing were reported previously\cite{HU2023168773}. Fig.~\ref{fig:t0-layout} is included here to identify the detector components relevant to the present readout study.

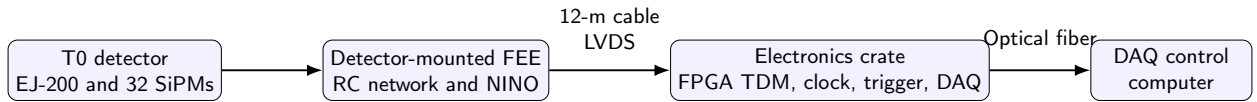
\begin{figure}
\centering
\resizebox{\linewidth}{!}{%
\begin{tikzpicture}[
  node distance=10mm,
  >=Latex,
  block/.style={
    draw,
    rounded corners,
    align=center,
    minimum height=9mm,
    minimum width=24mm,
    fill=blue!5
  },
  link/.style={->,thick}
]
\node[block] (detector) {T0 detector\\EJ-200 and 32 SiPMs};
\node[block,right=15mm of detector] (fee) {Detector-mounted FEE\\RC network and NINO};
\node[block,right=18mm of fee] (crate) {Electronics crate\\FPGA TDM, clock, trigger, DAQ};
\node[block,right=15mm of crate] (pc) {DAQ control\\computer};

\draw[link] (detector) -- (fee);
\draw[link] (fee) -- node[midway,above=2mm,align=center]{12-m cable\\LVDS} (crate);
\draw[link] (crate) -- node[midway,above=2mm,align=center]{Optical fiber} (pc);

\end{tikzpicture}}
\caption{Readout path from the T0 detector to the data-acquisition computer.}
\label{fig:readout-chain}
\end{figure}

Each SiPM produces a single-ended current pulse, whereas the NINO ASIC was originally developed for differential MRPC signals\cite{Anghinolfi2004}. One NINO ASIC contains eight amplifier--discriminator channels with adjustable thresholds and LVDS outputs. The T0 front-end electronics (FEE) is mounted as part of the detector assembly, where the SiPM signals are conditioned, amplified, and discriminated by NINO. The NINO LVDS outputs are transmitted through 12~m of high-density coaxial cable to the FPGA-based time-digitization module (TDM) in a remote electronics crate. The same crate contains the clock, trigger, and DAQ boards required for data acquisition. The acquired data are sent through optical fiber to the computer used to control data taking. Fig.~\ref{fig:readout-chain} summarizes this readout path.

The representative raw SiPM waveform rises in about 1.5~ns, carries a charge of several tens of picocoulombs, and has a tail extending over several hundred nanoseconds. Closely spaced pulses therefore overlap well before the intrinsic NINO processing rate is reached. Large input amplitudes can also induce crosstalk in adjacent NINO channels. The coupling stage must consequently shorten the tail and limit the input amplitude without adding a large or power-hungry active circuit.

\begin{figure}
\centering
\begin{tikzpicture}[>=Latex,thick]
  \node[draw,minimum width=16mm,minimum height=10mm,fill=yellow!10] (sipm) at (0,0) {SiPM};
  \coordinate (node) at (1.6,0);
  \draw (sipm.east) -- (node);
  \draw (node) -- (2.15,0);
  \draw (2.15,-0.33) -- (2.15,0.33);
  \draw (2.42,-0.33) -- (2.42,0.33);
  \draw (2.42,0) -- (3.2,0);
  \node[above] at (2.28,0.35) {$C_{\mathrm{s}}=50$ pF};
  \node[draw,minimum width=12mm,minimum height=7mm,fill=green!8] (res) at (1.6,-1.0) {$R_{\mathrm{p}}=100\,\Omega$};
  \draw (node) -- (res.north);
  \draw (res.south) -- (1.6,-1.65);
  \draw (1.25,-1.65) -- (1.95,-1.65);
  \draw (1.35,-1.80) -- (1.85,-1.80);
  \draw (1.47,-1.95) -- (1.73,-1.95);
  \node[draw,rounded corners,align=center,minimum width=22mm,minimum height=12mm,fill=blue!8] (nino) at (4.45,0) {NINO\\input};
  \draw[->] (3.2,0) -- (nino.west);
  \draw[->] (nino.east) -- ++(1.1,0) node[right]{LVDS};
\end{tikzpicture}
\caption{Simplified single-channel RC coupling network between a SiPM and the NINO input. The complete FEE also contains input protection and a matched reference branch.}
\label{fig:rc-network}
\end{figure}
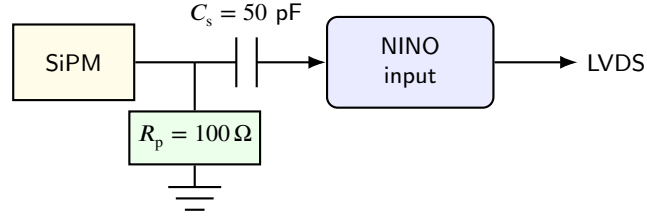

A passive differentiation and attenuation network was therefore placed directly before each NINO input. Fig.~\ref{fig:rc-network} shows the simplified single-channel topology: the series capacitor blocks the long low-frequency component and shortens the pulse tail, whereas the shunt resistor provides attenuation and a discharge path. Clamp protection and the matched reference branch used on the physical FEE are omitted from the simplified drawing. The network was simulated with Keysight Advanced Design System (ADS)\cite{ADS}. RC scans were then performed on the bench. The design objective was the shortest fall time compatible with a useful signal amplitude and reduced adjacent-channel crosstalk.

\begin{figure}
\centering
\includegraphics[width=0.75\linewidth]{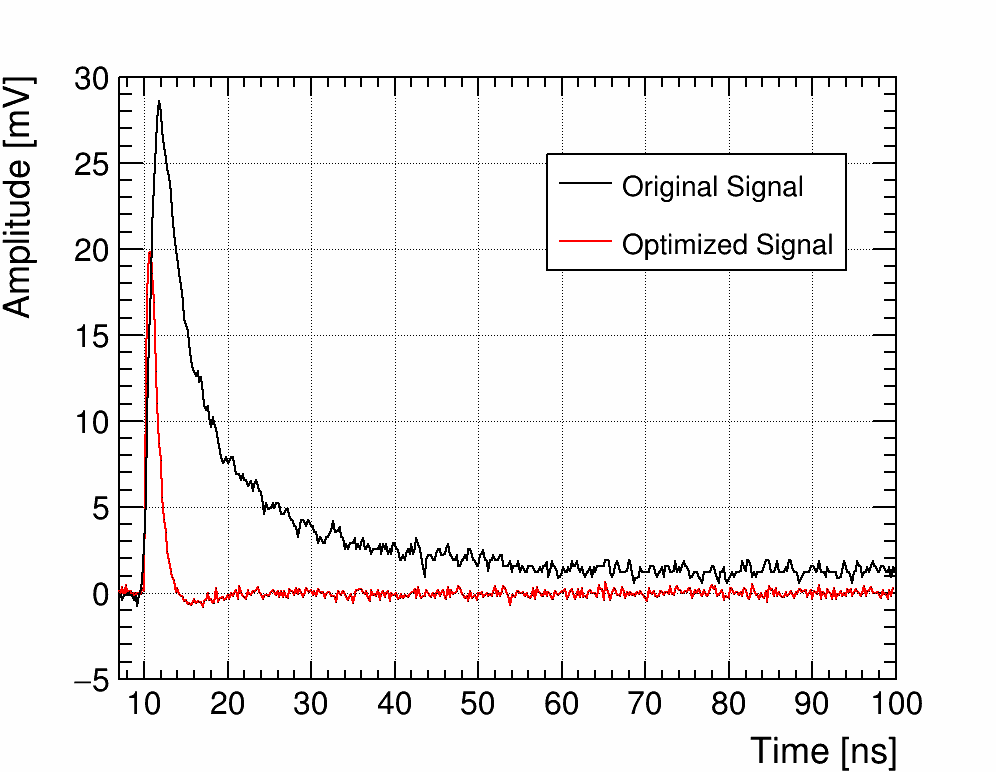}
\caption{Representative SiPM voltage waveforms before (black) and after (red) the 50-pF/100-$\Omega$ network. The network reduces the peak amplitude and suppresses the long pulse tail.}
\label{fig:shaped-waveform}
\end{figure}

The detailed optimization of the NINO input circuit is reported separately. Only the final configuration relevant to the present beam measurement is summarized here. In the end, a 50-pF series capacitor and a 100-$\Omega$ shunt resistor were selected. For the representative pulse in Fig.~\ref{fig:shaped-waveform}, the peak amplitude is reduced from about 28.5 to 19.5~mV (approximately 70\%), while the measured fall-time parameter is shortened to approximately 2.15~ns. The black waveform retains a long positive tail after its maximum, whereas the red waveform returns rapidly to the baseline. This comparison demonstrates the intended reduction of pulse overlap. It does not by itself establish an absolute counting efficiency at 100~MHz. So, the end-to-end rate test and its experimental scope are described next.

\section{Laboratory tests}
\label{sec3}

The installed light source was a Hamamatsu PLP-10 picosecond light-pulser system using a C10196 controller\cite{HamamatsuC10196}. The controller was operated with its internal trigger, for which 100~MHz is the highest selectable repetition frequency. Optical pulses were coupled through an FC interface to the lower part of the T0 detector. The light illuminated the scintillator rather than an individual SiPM, so scintillation light propagated to the SiPM arrays at both ends before electronic processing.

The NINO LVDS outputs were carried through the same 12-m link and TDM used in the beam test. In the DAQ beam-diagnostics mode, $M_{\mathrm{L}}$ and $M_{\mathrm{R}}$ were defined for each candidate event as the numbers of fired SiPM channels among the 16 channels at the left and right ends of the scintillator, respectively. A candidate event was recorded as one T0 hit only when both $M_{\mathrm{L}}\geq10$ and $M_{\mathrm{R}}\geq10$ were satisfied; events that did not satisfy this two-sided requirement were rejected and were not recorded. Consequently, every recorded entry in this mode represents one accepted T0 hit rather than an individual SiPM-channel signal. The same DAQ mode and multiplicity requirement were used for both the laboratory and in-beam measurements.

\begin{figure}
\begin{minipage}[t]{0.49\linewidth}
\centering
\includegraphics*[width=1\textwidth]{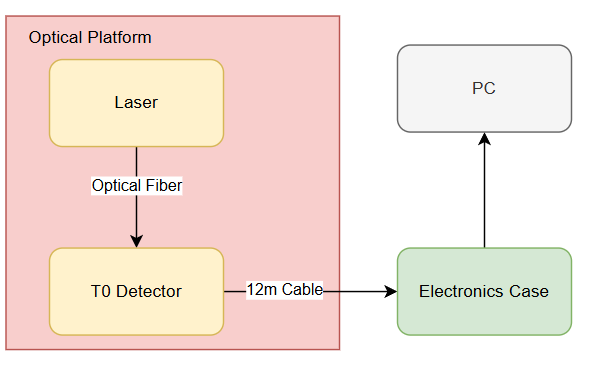}
\centerline{(a)}
\end{minipage}
\hfill
\begin{minipage}[t]{0.49\linewidth}
\centering
\includegraphics*[width=1\textwidth]{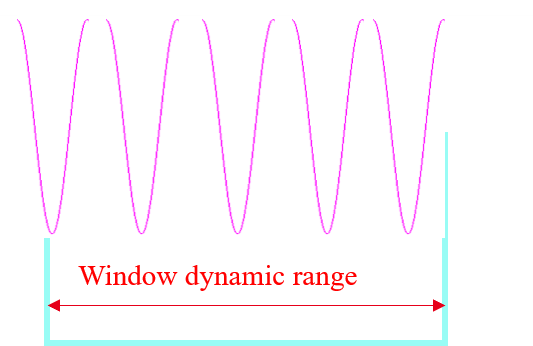}
\centerline{(b)}
\end{minipage}
\caption{Laser test and rate-estimation method. (a) End-to-end test arrangement. (b) Schematic programmable counting window; the drawn pulse train is illustrative.}
\label{fig4}
\end{figure}

For the 100-MHz measurement, NINO pulse stretching was disabled. The T0 hits recorded by the DAQ were analyzed offline. If $N$ recorded T0 hits occurred within a time window of duration $\Delta t$, the corresponding T0 hit rate was $R_{\mathrm{T0}}=N/\Delta t$. Fig.~\ref{fig4}(a) shows the end-to-end arrangement. The FEE is mounted on the T0 detector and is therefore not part of the ``electronics case.'' In this diagram, the ``electronics case'' denotes the remote crate containing the FPGA-based TDM and the clock, trigger, and DAQ boards. The crate is connected by optical fiber to the computer used to control data acquisition. Fig.~\ref{fig4}(b) illustrates the counting window used in the offline rate analysis; it is a schematic rather than a measured waveform.

\begin{figure}
\centering
\includegraphics[width=0.75\linewidth]{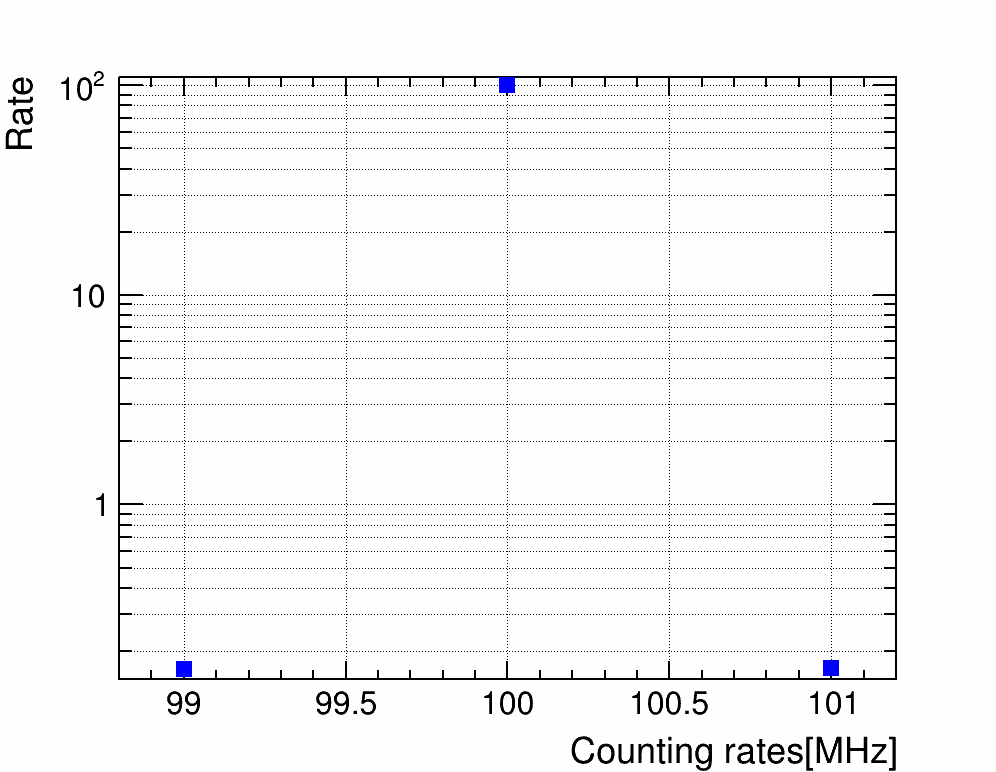}
\caption{Relative occurrence of T0 hit-rate estimates for the C10196 internal trigger set to 100~MHz. The dominant bin is normalized to 100.}
\label{fig:laser-100mhz}
\end{figure}

Fig.~\ref{fig:laser-100mhz} shows the T0 hit-rate estimates obtained with the C10196 internal trigger set to 100~MHz. The horizontal coordinate is the value ($N/\Delta t$) calculated for each offline analysis window, while the vertical coordinate is the normalized relative occurrence of that value, displayed on a logarithmic scale. Nearly all windows give a T0 hit rate of 100~MHz. The much smaller 99- and 101-MHz entries are consistent with finite-window boundary variations because the laser trigger is not phase locked to the analysis-window boundaries. The 100-MHz setting is the upper repetition frequency of the C10196 internal trigger.

This laboratory test verifies the high-rate capability of the T0 readout chain. With the C10196 internal trigger operated at its maximum repetition frequency of 100~MHz, the reconstructed T0 hit rate is concentrated at 100~MHz, demonstrating that the detector and readout electronics can reliably transmit, discriminate, and record signals at this rate. The result confirms an effective counting capability of 100~MHz under the laboratory test conditions. The in-beam observation in Section~\ref{sec4} provides a separate check with heavy-ion signals at lower instantaneous rates.

\section{Beam diagnostics}
\label{sec4}

\subsection{Experimental setup and scope of the beam measurement}
\label{sec:beam-setup}

\begin{figure}
\centering
\includegraphics[width=0.75\linewidth]{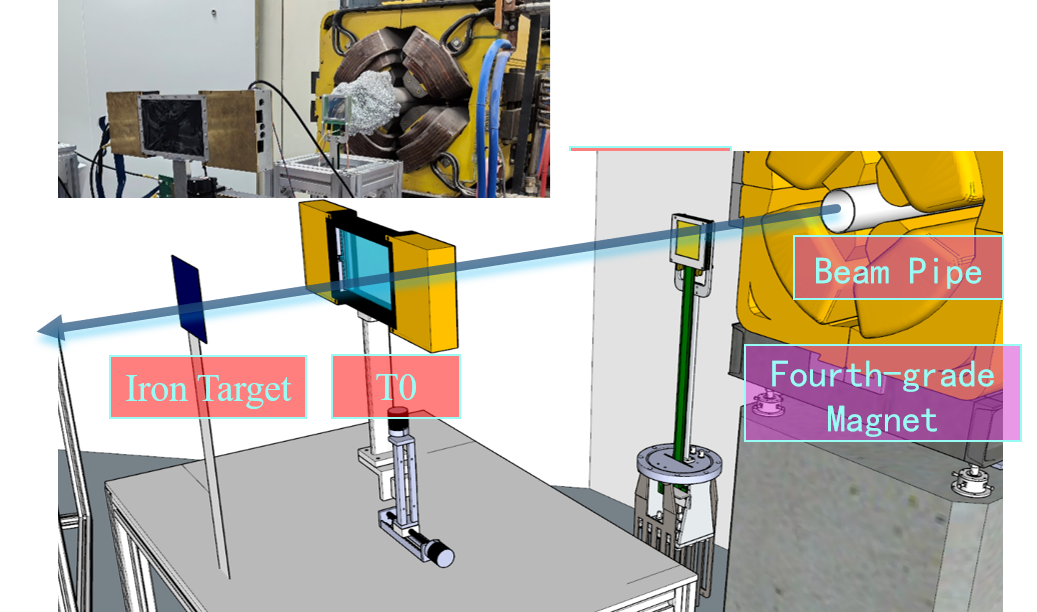}
\caption{Position of T0 in the CEE beam line. The beam travels from right to left through the final quadrupole and T0 before reaching the iron target; the inset shows the installed region.}
\label{fig:beam-setup}
\end{figure}

The T0 system participated in CEE beam tests at the Institute of Modern Physics in June 2024 and January 2025. The spill-structure measurements and accelerator-parameter scans presented in this work were acquired during the June 2024 beam test, which used a 350-MeV/u Kr beam. Several CEE detector subsystems participated in this beam test, including T0, the time-of-flight detectors, the TPC, and the MWDCs. In Fig.~\ref{fig:beam-setup}, the beam travels from right to left: it exits the beam pipe, passes the final quadrupole and T0, and then reaches the fixed iron target. T0 therefore samples the incident beam before interactions in the target. The accelerator cycle was approximately 30~s and contained one slow-extracted spill lasting about 3--4~s.

For the beam-structure measurements, the DAQ beam-diagnostics mode applied the same two-sided multiplicity requirement defined in Section~\ref{sec3} before recording each T0 hit. The slow-control monitoring system accumulated the recorded T0 hits over successive time windows and displayed the resulting T0 hit rate online, providing the time-dependent spill profile during beam commissioning. The displayed profiles were also recorded for subsequent offline extraction of the rise, plateau, and fall durations. The same discriminator-threshold configuration and DAQ event-selection criteria were retained when different accelerator settings were compared.

The beam data also demonstrate the need for a rate-capable electronic chain. During the June run, the T0 detector recorded as many as 20 hits in a 1-$\mu$s interval, corresponding to a T0 hit rate of 20~MHz in that interval. This value is 20 times the nominal 1-MHz T0 hit-rate target for CEE operation. Such short high-rate clusters within the extraction period would increase pile-up in a readout with a long pulse tail.

\begin{figure}
\begin{minipage}[t]{0.49\linewidth}
\centering
\includegraphics*[width=1\textwidth]{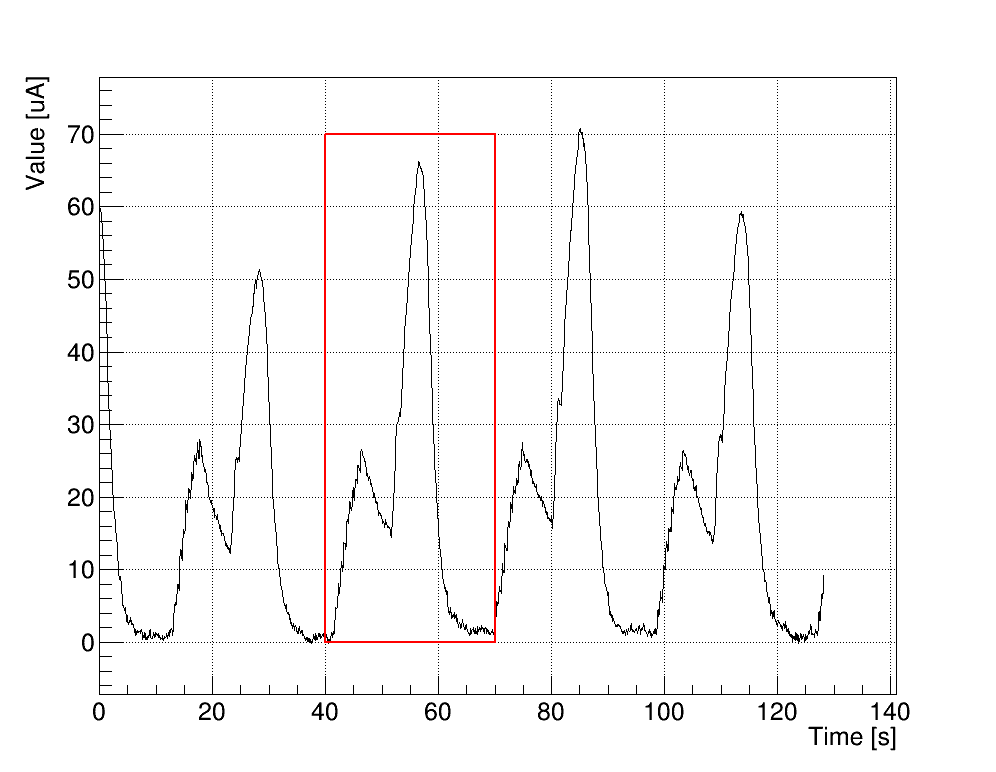}
\centerline{(a)}
\end{minipage}
\hfill
\begin{minipage}[t]{0.49\linewidth}
\centering
\includegraphics*[width=1\textwidth]{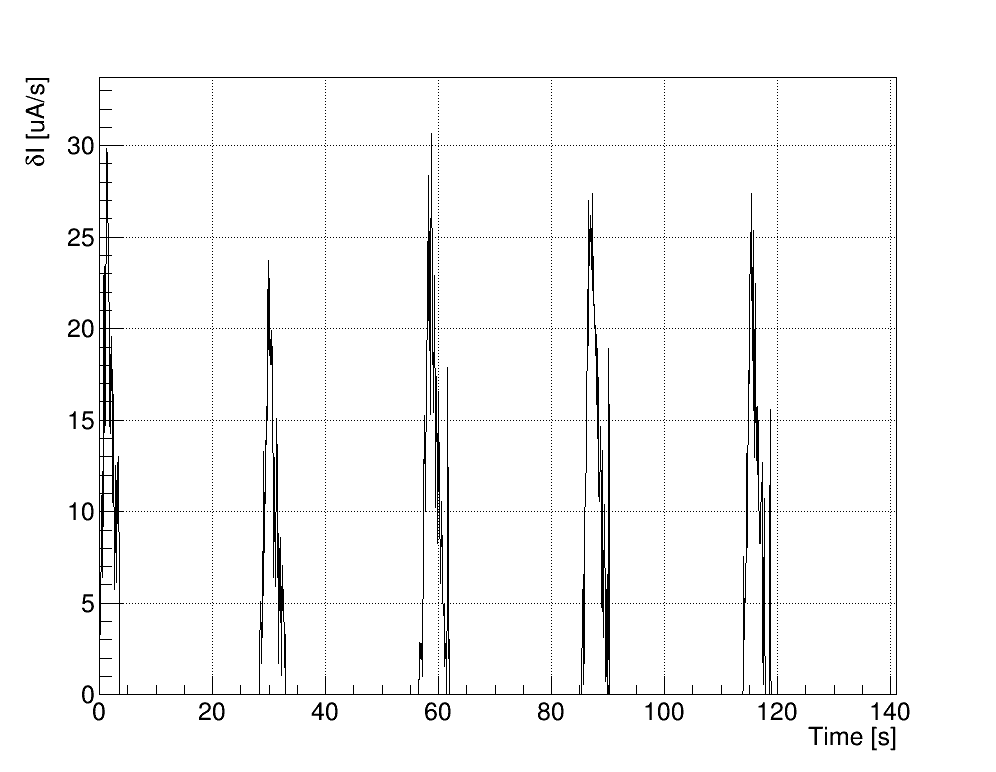}
\centerline{(b)}
\end{minipage}
\caption{Accelerator-current context for slow extraction. (a) Ring-current monitor over four machine cycles; the red box selects one cycle. (b) Magnitude of the ring-current decrease rate during extraction.}
\label{fig:ring-current}
\end{figure}

Fig.~\ref{fig:ring-current}(a) displays the accelerator ring-current monitor over four repeated machine cycles. The red rectangle marks the cycle between approximately 40 and 70~s that is used as an example. Within each cycle, the first, lower increase is associated with injection and capture, the second, higher peak follows acceleration, and the subsequent decrease occurs as the stored beam is slowly extracted. Fig.~\ref{fig:ring-current}(b) shows the positive magnitude of the current-decrease rate during the falling part of each cycle. If extraction were uniform, this rate proxy would be approximately constant during the spill; instead, it is concentrated in narrow, strongly modulated groups.

\begin{figure}
\centering
\includegraphics[width=0.78\linewidth]{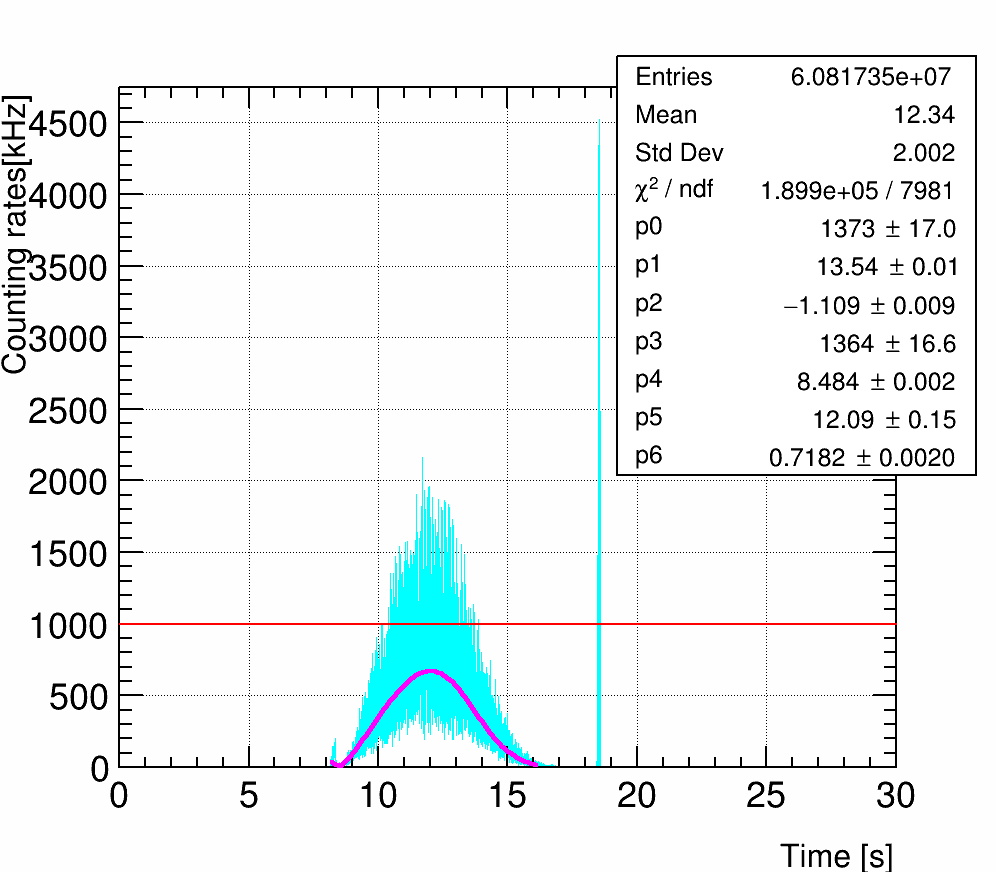}
\caption{Initial spill measured with T0. Cyan points are T0 hit rates, the magenta line is the empirical representation, and the red line marks the nominal 1-MHz target.}
\label{fig:initial-spill}
\end{figure}

During the June 2024 beam test, the T0 detector was operated in the same beam-diagnostics mode before physics data taking and after each extraction-setting change. Fig.~\ref{fig:initial-spill} presents the time-dependent T0 hit rate for an initial spill from this test. Cyan are the measured window-by-window rates, the magenta curve is the empirical smoothing function introduced below, and the red horizontal line marks the nominal 1-MHz T0 hit-rate target. The measured envelope is rounded and peaked rather than flat, and the isolated feature near 18.5~s is a terminal extraction spike. The temporal correspondence with the modulated current-decrease signal in Fig.~\ref{fig:ring-current}(b) supports an accelerator origin for the spill structure, although a single spill cannot identify the source of every fluctuation.

CEE operation seeks an approximately constant extracted-beam profile near the nominal 1-MHz T0 hit-rate target while avoiding short intervals above the approximately 4-MHz operating limit used for the TPC. The reconstructed T0 hit rate provides a direct measurement of the spill structure.

\subsection{Empirical spill-structure metrics}
\label{sec:spill-metrics}

For the present commissioning comparison, a desirable extracted-beam profile has three features: a short rise to the operating rate, a long interval near that rate, and a short terminal fall without a large end spike. Three descriptive durations are therefore used: rise, plateau, and fall. Together, they provide a compact description of the measured spill structure.

Because individual bins contain accelerator-induced fluctuations, an empirical composite curve is used only to obtain reproducible threshold crossings. It is the sum of a logistic component and a super-Gaussian component,

\begin{equation}\label{eq:spill-profile}
f(x) = \frac{L_0}{1 + e^{-k_0(x - x_0)}} + L_2 e^{-\left( \frac{(x - x_2)^2}{r^2} \right)^P}
\end{equation}

where $L_0$ and $L_2$ are the amplitudes of the logistic and super-Gaussian terms, respectively; $x_0$ is the logistic midpoint; $k_0$ controls the direction and steepness of that edge; $x_2$ is the centre of the super-Gaussian; $r$ is its width scale; and $P$ controls the flatness of its top. The function is not presented as a beam-dynamics model. It is an empirical smoothing representation that accommodates a broad spill envelope together with an asymmetric edge.

The $p0$--$p6$ values printed are the seven implementation parameters of this empirical curve; they are not treated as separate beam-quality observables. The red horizontal line in Figs.~\ref{fig:initial-spill} and \ref{fig:optimized-spill} denotes the nominal 1-MHz T0 hit-rate target for CEE operation. It is provided for visual comparison and is not the threshold used to calculate the three durations.

After subtraction of the pre-spill baseline, the empirical fitted curve is normalized to its maximum value. The rise duration is the interval between the first 5\% and 95\% crossings. The plateau duration is the interval from the rising 95\% crossing to the falling 95\% crossing, and the fall duration is the interval between the subsequent 95\% and 5\% crossings. Isolated narrow spikes are not used to define the normalization level. The same procedure and thresholds are applied to every setting. The empirical function does not reproduce all bin-to-bin variations in the measured spill profile, resulting in large reduced$ \chi^2$ values. It is therefore used only to determine the descriptive threshold-crossing times  Consequently, the curve is used for descriptive threshold extraction only. The initial and optimized spill profiles are shown as representative examples, while the accelerator-setting scans are presented through the extracted rise, plateau, and fall durations.

\subsection{Comparison of accelerator settings}

During the June 2024 spill-structure study, the accelerator team varied the platform RF voltage, maximum kick amplitude, RF-KO central frequency, sextupole ramp, and feedback gain. T0 recorded the spill profile after each adjustment. The analysis is limited to comparing the measured profiles and does not assign a beam-dynamics mechanism to the observed changes. Each plot is a one-dimensional operational scan: the named variable was changed while the other settings were retained at the values used for that scan. After every change, T0 supplied an independent measurement of the spill structure.

To compare the T0 measurements consistently during accelerator tuning, long-term accelerator stability and an acceptable RF-system load were considered first. Among the settings that satisfied these constraints, the duration of the near-constant plateau was treated as the primary spill-quality metric. When several settings produced comparable plateau durations, the rise duration and the operating margin from a rapidly deteriorating region were used as secondary considerations. The fall duration was retained as a descriptive metric but was not the primary quantity used in the comparison.

\begin{figure}
\centering
\includegraphics[width=0.75\linewidth]{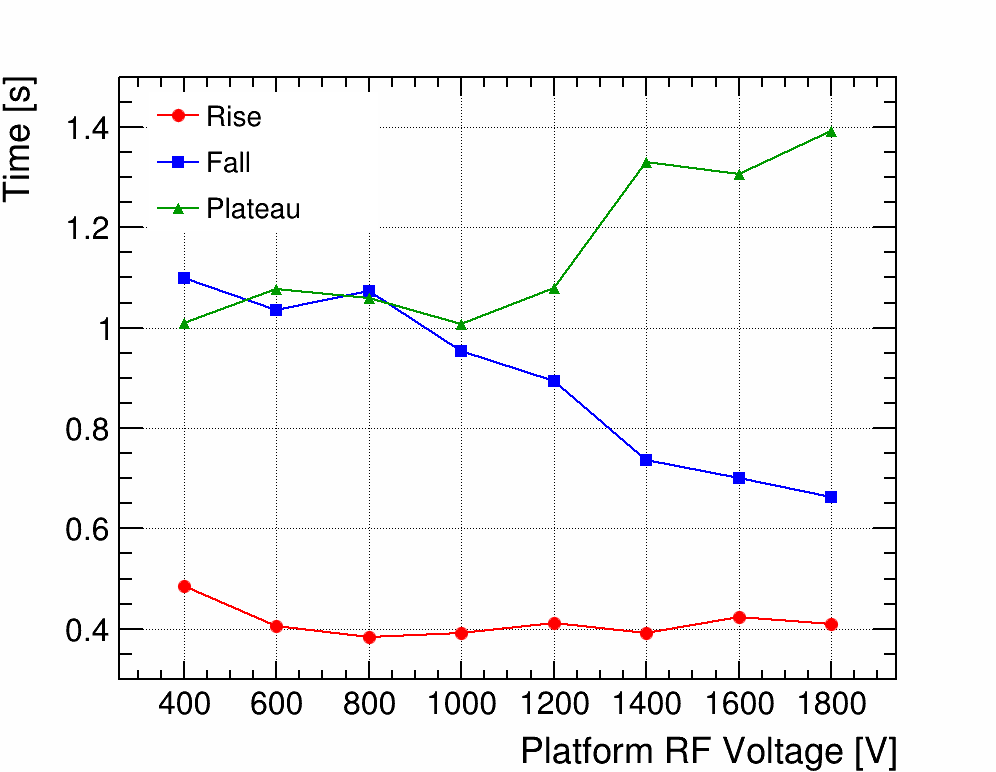}
\caption{Rise (red circles), plateau (green triangles), and fall (blue squares) durations obtained in the platform RF voltage scan.}
\label{fig:rf-voltage}
\end{figure}

Fig.~\ref{fig:rf-voltage} displays the three durations obtained during the platform RF voltage scan. The plateau generally becomes longer as the voltage increases, whereas the rise duration changes less and the fall duration decreases. However, the higher-voltage settings imposed an excessive load on the RF system and could not be maintained during extended operation. A platform RF voltage of 1200~V was therefore retained as the highest practical setting for long-term operation, rather than selecting the voltage that gave the numerically longest plateau in the short diagnostic scan.

\begin{figure}
\begin{minipage}[t]{0.49\linewidth}
\centering
\includegraphics*[width=1\textwidth]{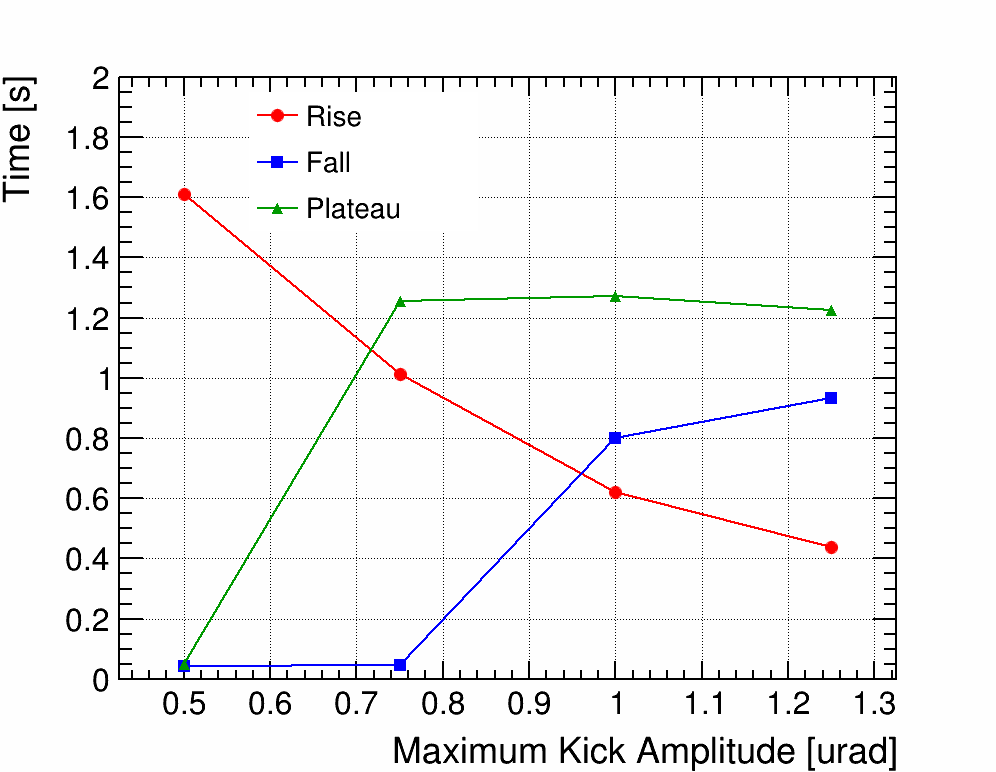}
\centerline{(a)}
\end{minipage}
\hfill
\begin{minipage}[t]{0.49\linewidth}
\centering
\includegraphics*[width=1\textwidth]{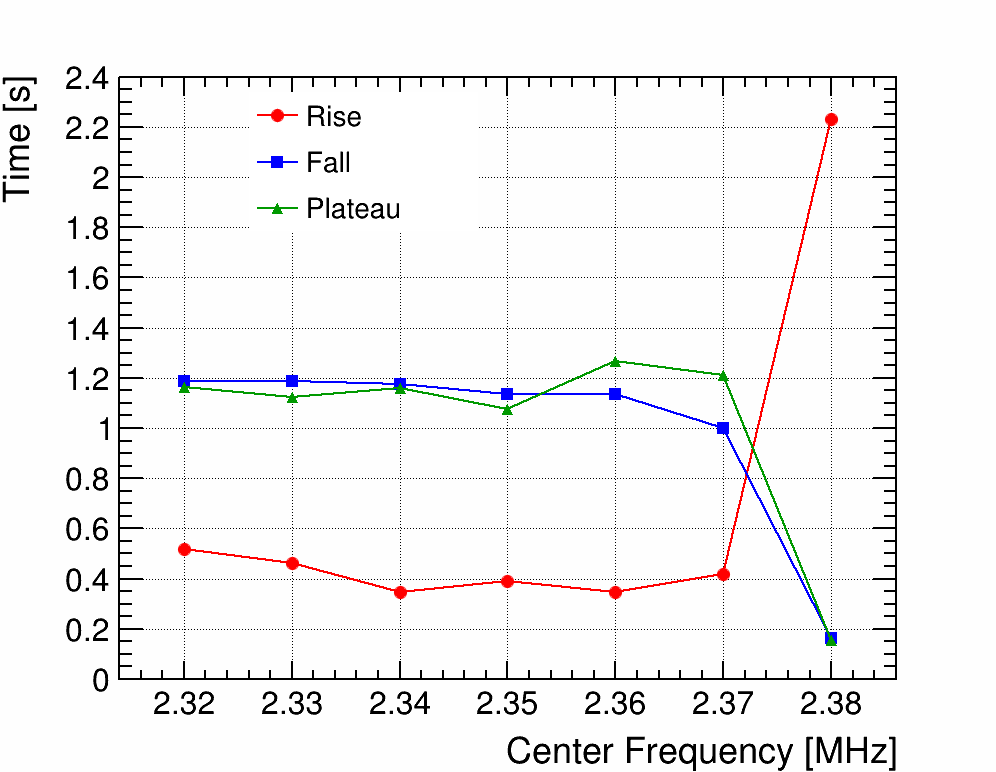}
\centerline{(b)}
\end{minipage}
\caption{Rise (red circles), plateau (green triangles), and fall (blue squares) durations from the RF-KO scans: (a) maximum kick amplitude and (b) central frequency.}
\label{fig:rfko-scans}
\end{figure}

Fig.~\ref{fig:rfko-scans}(a) shows the maximum kick amplitude scan. At 0.5~$\mu$rad the empirical curve has essentially no 95\%-level plateau and a long rise, and therefore does not provide the desired extraction envelope. Between 0.75 and 1.25~$\mu$rad the plateau remains close to 1.2~s and therefore changes little according to the primary selection metric. Within this group of settings with comparable plateaus, the rise becomes progressively shorter as the maximum kick amplitude increases, although the fall becomes longer. A maximum kick amplitude of 1.25~$\mu$rad was consequently retained because it preserved the developed plateau while giving the shortest rise among the tested points.

Fig.~\ref{fig:rfko-scans}(b) compares the T0 profiles recorded after the accelerator team changed the RF-KO central-frequency setting. The measured profiles are comparatively stable from 2.32 to 2.37~MHz, with similar plateau durations throughout most of this interval. After the setting was changed to 2.38~MHz, T0 measured a sharp decrease in the plateau and fall durations and a rise duration above 2~s. Because the present measurements do not include accelerator beam-spectrum information, no mechanism is assigned to this change. The T0 comparison therefore provided experimental support for using 2.35~MHz, near the middle of the stable interval, during subsequent accelerator tuning.

Two sextupole-ramp patterns provided by the accelerator team were tested and are denoted configurations A and B in Table~\ref{tab:sextupole}. The displayed $k_2$ values are reported as accelerator control-system set points. Configuration A gives a substantially longer plateau, whereas configuration B gives a shorter rise but a long fall. Because useful near-constant extraction time was the primary operational criterion, configuration A was retained.

\begin{table}
\centering
\small
\caption{Spill-profile durations for two sextupole-ramp configurations. Arrows give the initial and final $k_2$ control-system set points; 22s02, 23s01, and 23s02 are the accelerator magnet identifiers.}
\label{tab:sextupole}
\renewcommand{\arraystretch}{1.2}
\begin{tabular}{c p{6.3cm} ccc}
\toprule
Config. & Sextupole ramps & Rise (s) & Plateau (s) & Fall (s) \\
\midrule
A & 22s02: $0.4\rightarrow1.3$; 23s01: $0.4\rightarrow1.3$; 23s02: $0.4\rightarrow1.3$
& 0.614 & 1.018 & 1.137 \\
B & 22s02: $0.8\rightarrow2.0$; 23s01: $0.4\rightarrow1.5$; 23s02: $0.8\rightarrow2.0$
& 0.151 & 0.245 & 1.861 \\
\bottomrule
\end{tabular}
\end{table}

The feedback-gain scan is summarized in Table~\ref{tab:feedback}. The listed strength is the dimensionless gain setting in the accelerator control interface. The T0 detector independently records its result. Relative to zero feedback, a gain of 1.0 reduces the rise duration by 62\%, increases the plateau duration by 132\%, and reduces the fall duration by 86\%. A gain of 10.0 gives a slightly shorter fall, but its plateau is shorter and its rise is longer than at 1.0. Gain 1.0 was therefore selected because it produced the longest plateau among the tested settings while also maintaining short rise and fall durations.

\begin{table}
\centering
\caption{Spill-profile durations obtained in the feedback-gain scan. The gain is the dimensionless setting displayed by the accelerator control system.}
\label{tab:feedback}
\renewcommand{\arraystretch}{1.3}
\begin{tabular}{cccc}
\toprule
Feedback gain & Rise (s) & Plateau (s) & Fall (s) \\
\midrule
0.0  & 0.614 & 1.018 & 1.137 \\
1.0  & 0.233 & 2.360 & 0.163 \\
10.0 & 0.460 & 2.150 & 0.120 \\
\bottomrule
\end{tabular}
\end{table}

\begin{figure}
\centering
\includegraphics[width=0.75\linewidth]{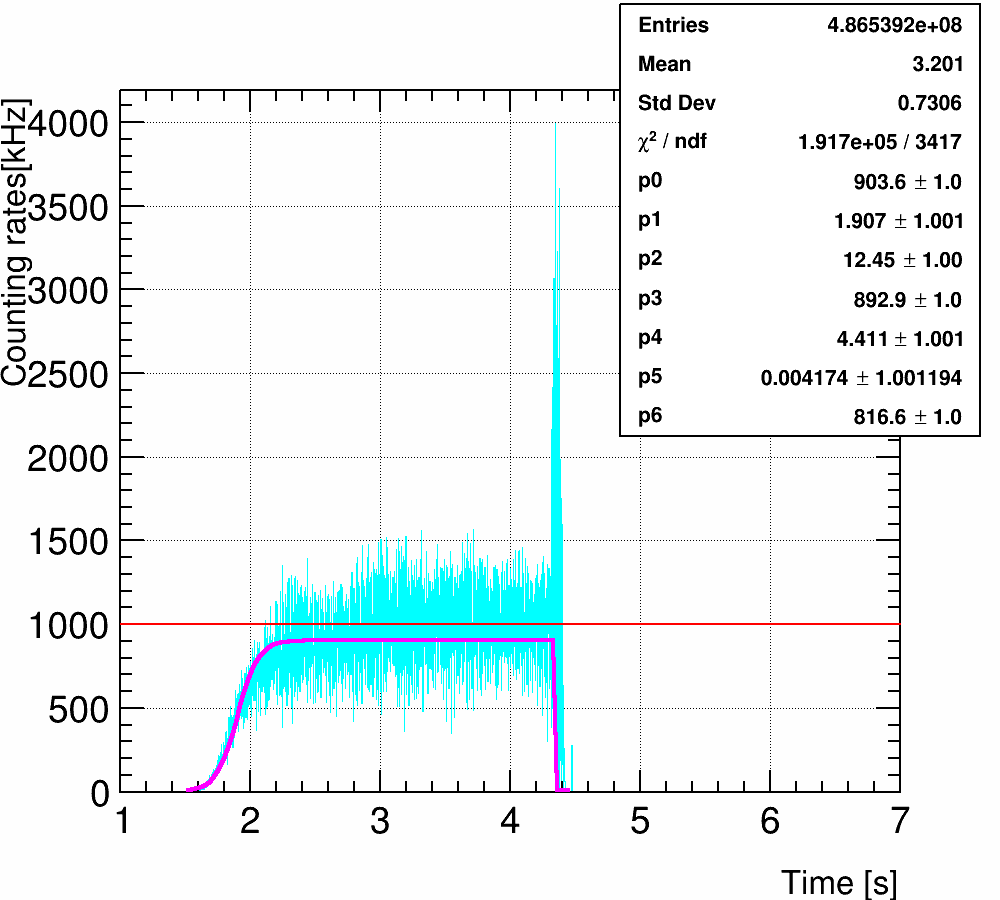}
\caption{Spill measured with the selected extraction settings. Cyan points are T0 hit rates, the magenta line is the empirical representation, and the red line marks the nominal 1-MHz target.}
\label{fig:optimized-spill}
\end{figure}

Combining the operational selections gives a platform RF voltage of 1200~V, a maximum kick amplitude of 1.25~$\mu$rad, a central RF-KO frequency of 2.35~MHz, sextupole configuration A, and feedback gain 1.0. In Fig.~\ref{fig:optimized-spill}, the magenta representation rises to approximately 0.9~MHz, remains nearly constant from about 2.2 to 4.3~s, and then falls rapidly. This is more rectangular than the rounded initial profile in Fig.~\ref{fig:initial-spill}. The individual cyan measurements nevertheless retain substantial bin-to-bin fluctuations, and a terminal spike reaches several megahertz. The result is therefore described as an improved spill profile rather than an ideal or uniform one.

\section{Summary}
\label{sec5}
This work adds a beam-monitoring function to the existing CEE T0 detector and applies it during accelerator tuning. The detector uses a 0.8-mm EJ-200 scintillator read out by 32 Hamamatsu S13360-3025PE SiPMs. Its readout incorporates a passive 50-pF/100-$\Omega$ shaping network to mitigate pulse overlap and make the long single-ended SiPM pulse more suitable for the NINO front end. For a representative pulse, the amplitude is reduced to about 70\% of its original value, while the fall time is shortened to approximately 2.15~ns. Using the Hamamatsu PLP-10 laser system with the C10196 controller and with NINO pulse stretching disabled, stable end-to-end digital output was observed up to the controller's 100-MHz internal-trigger limit, providing the readout-performance basis for the beam measurements.

The detector system was subsequently used during slow-extraction beam commissioning at HIRFL. The spill-structure results reported here were obtained in the June 2024 beam test, although the T0 system also participated in later CEE beam measurements. In the DAQ beam-diagnostics mode, the T0 hit-rate profiles were recorded after accelerator-parameter adjustments. The recorded profiles were subsequently analyzed offline.

The accelerator-setting scans conducted during commissioning, including changes to the platform RF voltage, maximum kick amplitude, RF-KO central frequency, sextupole ramp, and feedback gain, reveal clear differences in the measured spill profile. The spill structure was quantitatively characterized in terms of its rise, plateau, and fall durations. In the feedback-gain scan, for example, gain 1.0 changes these durations from 0.614, 1.018, and 1.137~s to 0.233, 2.360, and 0.163~s, respectively. The resulting spill exhibits a substantially extended plateau and much shorter rise and fall regions, giving a more nearly rectangular time profile than the initial spill, although a residual terminal spike remains.

The T0 detector provides an independent measurement of the beam time structure at the CEE target location after each adjustment of the accelerator settings. By extracting the rise, plateau, and fall durations of the spill, the effects of different accelerator settings on the slow-extracted beam structure can be compared. After taking accelerator stability and RF-system load into account, the plateau duration can serve as one of the primary metrics for evaluating the spill profile. With this additional capability, the existing T0 detector can serve not only as a start-time detector for event timing but also as an independent online beam-diagnostics tool, providing complementary information for accelerator tuning and beam-quality assessment without directly participating in the accelerator feedback loop.

More generally, this study demonstrates that the functionality of fast-timing detectors need not be limited to precise timing measurements. Fast-timing detectors already installed along the beam line or in the experimental area for event timing can also exploit their high time resolution and high-rate capability to monitor the temporal structure of slow-extracted beams. They can therefore serve as complementary diagnostic devices alongside dedicated beam-monitoring systems, providing an additional approach to beam-structure monitoring in other heavy-ion experiments using slow extraction.

\section*{Acknowledgments}

The authors thank the USTC high-energy physics group.
\printcredits

\bibliographystyle{unsrtnat}
\bibliography{references}

\end{document}